Title: Testing three models of cognitive stress effects: A psychopharmacological randomized controlled trial of acute stress and stress hormones across visual perception, response inhibition and cognitive flexibility

Authors:

*Lisa Weckesser, Institute of Clinical Psychology an Psychotherapy, Technische Universität Dresden, Dresden, Germany

Charlotte Grosskopf, Department of Psychiatry an Psychotherapy, Carl Gustav Carus University Hospital, Technische Universität Dresden, Dresden, Germany; Charlotte.Grosskopf@ukdd.de

Benjamin Weber, Department of Psychiatry, Sächsisches Krankenhaus Arnsdorf, Arnsdorf, Germany; Benjamin.Weber@skhar.sms.sachsen.de

Selen Soylu, Institute of Clinical Psychology an Psychotherapy, Technische Universität Dresden, Dresden, Germany

Tanja Endrass, Institute of Clinical Psychology an Psychotherapy, Technische Universität Dresden, Dresden, Germany

Robert Miller, Psychologische Hochschule Berlin, Berlin, Germany

**Abstract**: Acute stress alters cognitive performance, yet competing models make divergent predictions regarding the mechanisms, scope, and temporal dynamics of these effects. This large-scale randomized controlled trial tested predications from three influential stress-effect models using a broad cognitive task battery embedded within a psychopharmacological stress paradigm. Across 606 testing sessions, 303 healthy male participants completed both the Maastricht Acute Stress Test (MAST) and its non-stress control condition. To independently manipulate acute stress and stress hormone availability, participants were additionally randomized to receive atomoxetine (40 mg; to prolong norepinephrine availability), hydrocortisone (10 mg; to increase cortisol availability), or placebo. Cognitive performance was assessed over 80-minutes (post-stress) using tasks targeting visual perception (rapid serial visual presentation), response inhibition (stop-signal), and cognitive flexibility (dual and switch tasks). MAST exposure selectively impaired response inhibition, reflected in shorter stop-signal delays ($X^2(2)=34.20$, $p<0.0001$), lower probabilities of successful stopping ($X^2(2)=47.429$, $p<0.0001$) and prolonged stop-signal reaction times ($X^2(2)=147.01$, $p<0.0001$), particularly during later testing phases (40-80 minutes post-stress). MAST exposure did not affect visual perception or task-switching performance but buffered time-related declines in processing efficiency ($X^2(2)=16.52$, $p<0.0001$) at the expense of task prioritization ($X^2(2)=7.16$, $p=0.0279$) in the dual task. Pharmacological manipulation of norepinephrine or cortisol availability was effective but did not moderate cognitive stress effects. Overall, this pattern of task-specific impairment alongside stabilized processing efficiency cannot be fully explained by any tested model, highlighting the need to refine existing models and adopt more integrative approaches to advance our mechanistic understanding of cognitive stress-effects in laboratory and real-world contexts.

**Keywords (5)**: Stress, Cognition, Cortisol, Atomoxetine, MAST;

**Number of words**: 9015 or without figures and legends/title, abstract and references: 7676

# 1 Introduction

Similar to high-performing athletes who may experience temporary performance gains before over-training or burnout, acute stress exerts both positive and negative effects on cognitive performance in the laboratory with respect to stressor type, timing, and task(Lemyre et al., 2007; Martinent et al., 2020; Olsson et al., 2024). A major meta-analysis suggests that acute stress impairs working memory (the ability to retain or retrieve information no longer available) and cognitive flexibility performance (the ability to adapt responses to changing contextual information(Shields et al., 2016)). Conversely, stress appears to enhance response inhibition performance (the ability to suppress no longer relevant or maladaptive responses(Shields et al., 2016)), whereas its effect on visual-sensory processing, potentially constituting a building block for contextual information processing and thus for working memory, flexible decision-making and response inhibition, has hardly been the focus of research (Schwabe & Wolf, 2010; Shields et al., 2016; Weckesser et al., 2016). To explain such heterogeneous effects of acute stress on cognitive performance, different stress effect models were proposed (Arnsten, 2019; Schwabe et al., 2022; van Oort et al., 2017). Four influential classes of these models, visualised in **Figure 1**, differ in their predictions about which cognitive processes and related task performances are most affected by acute stress and which neurophysiological mediator is driving these effects, focussing on the stress hormones norepinephrine, NE and cortisol, CORT.

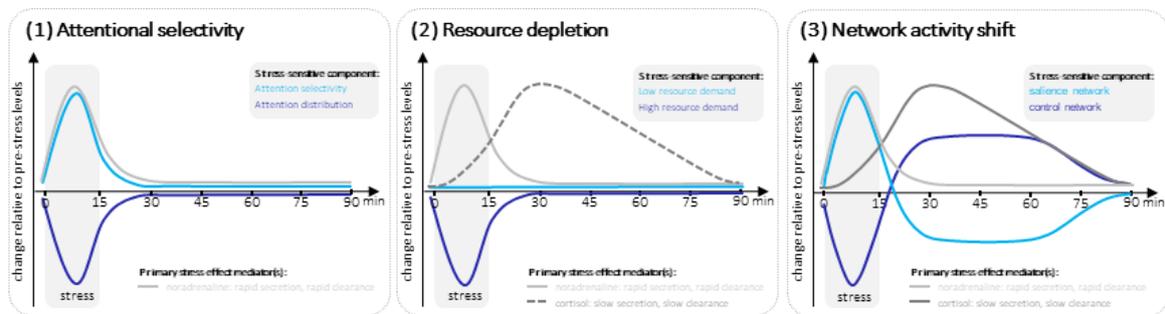

**Figure 1.** Schematic illustration of three influential classes of cognitive stress-effect models and their predictions regarding stress-induced changes in cognitive processing as a function of norepinephrine (NE) and cortisol (CORT) availability. **(A) Attentional selectivity models** propose that acute stress induces a relevance-based narrowing of attention, favouring salient or task-relevant information at the expense of peripheral or less relevant inputs, primarily driven by increased noradrenergic activity (Berridge & Spencer, 2016; Chajut & Algom, 2003; Wang et al., 2022). **(B) Resource depletion models** assume that coping with acute stress consumes finite cognitive resources, thereby sparing low-demand or automatic processing while impairing high-demand executive control processes. These effects are attributed mainly to NE and may be amplified by CORT (Arnsten, 2019; Hockey, 1997; van Marle et al., 2009). **(C) Network activity shift models** posit a dynamic rebalancing of large-scale brain networks under stress, with an initial NE-driven shift toward salience-network dominance followed by a CORT-related re-engagement of control-network activity over time (Hermans et al., 2014; van Oort et al., 2017).

As visualised in **Figure 1**, the three model classes attribute stress effects on cognitive processing either to (1) attentional selectivity, (2) resource depletion, or (3) stress hormone pharmacokinetics and the stress hormone induced activity shifts between neuronal networks. Beside their differences, however, they share the assumption with model class (3) that CORT and/or NE are primary stress effect mediators. Upon stress onset, the sympatho-adreno-medullary system rapidly increases NE secretion into the coeruleo-cortical-NE system (Engert et al., 2011; Silverberg et al., 1978). NE levels peak within minutes and returns to pre-stress levels within 15 minutes (Engert et al., 2011; Silverberg et al., 1978). NE binds to adrenoceptors (ARs), particularly α2-autoreceptors, which are densely expressed in frontal and occipital brain regions (Jones & Palacios, 1991; Joyce et al., 1992). Since NE seems to reduce its own release in these regions, the (1) stress hormone pharmacokinetics model presumes that stress (NE) initially impair cognitive flexibility, response inhibition and visual-sensory processing (cf. **Figure 2**; Alexander et al., 2007; Berridge & Spencer, 2016; Chang et al., 2020; Gathmann et al., 2014; Meier & Schwabe, 2024; O'Hanlon, 1964; van Marle et al., 2009). Meanwhile, the hypothalamic-pituitary-adrenal axis stimulates CORT secretion, which starts to enter the bloodstream with a delay of about 15 minutes (Engert et al., 2011; Miller et al., 2017; Vogel et al., 2016). Under stress (i.e., high CORT availability), excess CORT primarily binds to its low-affinity glucocorticoid receptors (GRs), which are particularly densely expressed in occipital brain regions (as well as hippocampal brain regions; Perlman et al., 2007; Vogel et al., 2016; Webster et al., 2002). Since CORT seems to enhance occipital activity, the (1) stress hormone pharmacokinetics model presumes that stress (CORT) should improve visual-sensory processing, especially later in time (Muehlhan et al., 2020; Schwabe & Wolf, 2010; van Marle et al., 2009; Weckesser et al., 2016).

Model class (3) bases its stress-effect predictions on the time-dependent availability of the stress hormones NE and CORT (Hermans et al., 2014; Joëls, 2018). In the (1) stress hormone pharmacokinetic models, stress-induced performance changes are driven solely by the distinct time courses of NE and CORT, and their differential effects on brain regions that are closely related with task processing. As stress-induced hormone availability declines, activity in these regions, and consequently, related task performance, should gradually return to levels observed under non-stress control conditions. In contrast, the (4) network activity shift models link task performance to distributed neuronal networks (i.e., the salience and the control network) spanning multiple brain regions with varying stress hormone receptor densities. They suggest that changes in stress hormone availability shift the relative dominance of the salience network toward the control network activity, leading to a time-dependent improvement or impairment of task performance (depending on which network dominates) under stress compared to non-stress control conditions (Henze, 2025; Hermans et al., 2014; Schwabe, 2017).

Model classes (2) and (3), both rooted in cognitive psychology, share the basic assumption that coping with stress depletes available attentional or cognitive processing resources (Chajut & Algom, 2003; Hockey, 1997). However, they differ in how they explain the consequences of this depletion. The (2) attentional selectivity model suggests that stress-induced depletion is compensated by a strategic shift, which increases selectivity in information processing by focussing attention to contextually relevant information (effectively filtering our irrelevant information; Chajut & Algom, 2003; Gathmann et al., 2014). This regulatory strategy optimizes the use of limited processing resources, helps to maintain performance under stress and is linked to stress-induced NE secretion, which may increase neuronal selectivity and/or mobilize additional processing resources (Berridge & Spencer, 2016; Chajut & Algom, 2003; Hockey, 1997). In contrast, the (3) resource depletion model presumes that stress depletes processing resources without any compensation to maintain performance (Chajut & Algom, 2003). As a result, the remaining resources are only sufficient for processing highly salient or easily accessible information, regardless of its contextual relevance, because saliency captures processing resources without the need for effortful resource allocation. Van Marle (2009; p.649) describes this as a state of indiscriminate hypervigilance, in which stress prioritizes the detection of salient (potentially threatening) information to reduce the risk of missing critical information, even at the expense of overall performance.

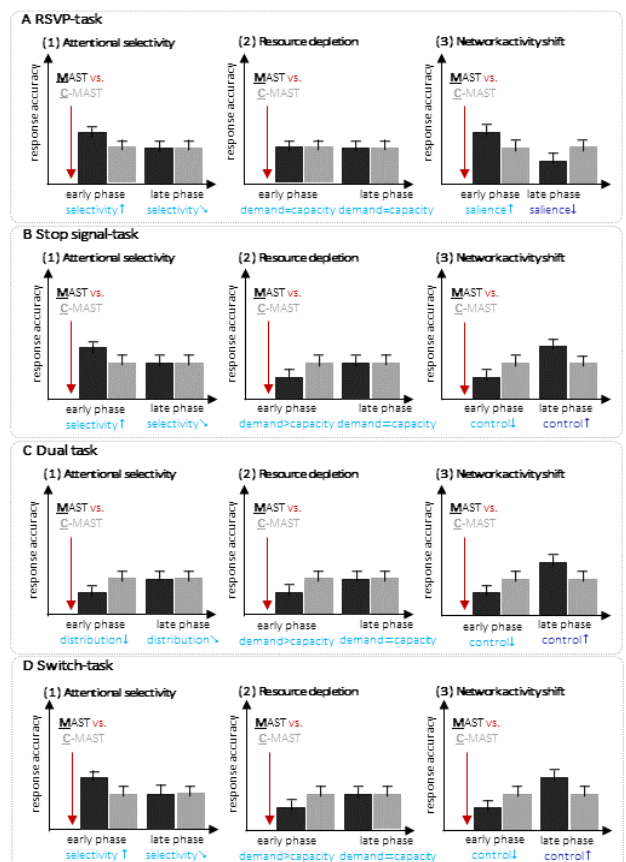

Figure 2. Schematic illustration of the predicted response accuracy patterns for the three model classes across the different cognitive tasks, as a function of early (norepinephrine NE-dominated; 0-40 minutes after stress/control test offset) and late (cortisol CORT-dominated; 40-80 minutes after stress/control test offset) phases of acute stress responses: The rapid serial visual presentation (**RSVP**) **task** indexes low-demand, occipitally mediated visual–sensory pro-cessing. The **stop-signal task** indexes high-demand, frontally mediated response inhibition (inhibitory control). The dual and switch-task index high-demand, frontally mediated cognitive flexibility, requiring either the prioritization of one of two concurrently presented tasks based on presentation order (**dual task**) or switching between tasks based on a preceding visual cue (**switch-task**). The **attentional selectivity model** (first [1] column in **A** to **D**) predicts that stress and NE enhance selective attention to task-specific cues and stop signals, thereby improving accuracy in the RSVP, stop-signal and switch tasks. However, stress should impair attention distribution across two tasks, leading to impaired dual-task accuracy (Alexander et al., 2007; Chajut & Algom, 2003; Ishizuka et al., 2007). The **resource depletion model** (second [2] column in **A** to **D**) predicts that stress and NE impair accuracy in all high-demand tasks while leaving RSVP task accuracy unaffected (Chajut & Algom, 2003; van Marle et al., 2009). The **network activity shift model** (third [3] column in **A** to **D**) predicts that under stress, the initial NE-driven shift toward the salience network should enhance RSVP task accuracy but impair accuracy in all three tasks requiring high control network activity. The delayed CORT-increase should reallocate activity back toward the control network, improving accuracy in the stop-signal, dual and switch tasks while impairing RSVP task accuracy (Hermans et al., 2014). Additional notes: Light blue shading represents NE-related effects, while medium blue shading indicates CORT-related effects. Atomoxetine (40mg) may enhance NE-related stress effects, while hydrocortisone (10mg) may enhance CORT-related stress effects (not illustrated here for clarity).

Proceeding from the assumptions of these cognitive processing models under stress to predict performance (specifically response accuracy) in four cognitive tasks, we obtained the performance predictions shown in **Figure 2**. To test these competing model predictions, we conducted a randomized factorial trial with 303 men aged 30 to 45 years. Participants were exposed to the Maastricht Acute Stress Test (MAST) and, one week later, to its control test (C-MAST; Quaedflieg et al., 2017; Smeets et al., 2012). Additionally, they received either 40 mg of atomoxetine (AX), 10 mg of hydrocortisone (HC), or placebos (PL), with AX and HC potentially enhancing the magnitude and duration of the suspected NE- or CORT-related stress effects (Chamberlain et al., 2009; Sauer et al., 2005; Weckesser et al., 2016). Following C-/MAST exposure and substance ingestion, participants completed twelve mini-blocks of two different two-alternative choice reaction tasks (in alternating order) to ensure a sufficient number of trials per task during the 60-min post-exposure testing period, with pre-exposure task practice (Zerbes & Schwabe, 2021).

## 2 Methods

### 2.1. Participants

A total of 303 healthy male participants were enrolled in this trial (mean age: 24.63 ± 4.35; age range: 18 - 35 years; mean BMI: 22.33 ± 2.21 kg/m$^2$). All participants reported having normal or corrected-to-normal vision, no history of psychiatric or chronic disorders, no regular use of medication or drugs, and smoking no more than ten cigarettes per day. Participant were recruited through the *Technische Universität Dresden*'s participant database, as well as student and staff directories, and local advertisements. To minimize interference from glucose and attenuate that of caffeine on CORT and NE measurements, participants were instructed to refrain from consuming anything except of water for at least 2 hours before their testing (Kirschbaum et al., 1997). This study received ethical approval from the local ethics committee (IRB00001473 and IORG0001076; dossier EK 492122017). All participants provided written informed consent before participation and were compensated 100€ for their participation upon completing the second of their two testing sessions.

### 2.2. Preregistration and modifications

This study was preregistered under ISRCTN65723653 (https://doi.org/10.1186/ISRCTN65723653). We deviated from the initial preregistration in three minor points: First we increased the upper age limit of our participants from 35 to 40 years to expand the participant pool and facilitate recruitment. Second, we (exploratively) analysed the number of trials completed per mini-task block and session as a tertiary outcome measure of motivation. Third, we enrol 303 participants instead of the initially planned 320 (or 324 when accounting for a 2% dropout rate) due to time constraints within the funding period.

## 2.3. Sample size

The required sample size for this trial was estimated *a priori* using GPower 3.1.9 (Faul et al., 2009). The probability of a family-wise error to incorrectly label an actually absent effect of stress as "significant" was set at α=5% (two-sided). Meanwhile, the probability of failing to detect the smallest effect of interest (SESOI (Lakens et al., 2018)) was set to β=20%, implying a statistical power of 80%.

Previous studies suggest that the effects of acute stress on response accuracy tend to be larger than those on response times, though both are relatively small(Shields et al., 2016). Reported effect sizes include ΔM(AC)/SD(AC) = 3%/10%, corresponding to a SESOI of d = 0.3 for stop-signal (Chamberlain et al., 2009; Shields et al., 2015, 2016), dual-, and switch-task accuracy in unselected source populations (Dierolf et al., 2016; Shields et al., 2016; Weckesser et al., 2016). To enhance statistical power, this trial primarily recruited participants from a homogeneous university population and employed a mixed-factor design (cf. **Figure 3**), which was assumed to reduce residual variance in response accuracy by approximately 20% (Jager et al., 2017). Proceeding from this assumption the SESOI effectively increases d = 0.39, so that Type II error rate was supposed to be lower than 20%. Consequently, this trial was designed to achieve a statistical power of at least 80% to detect SESOIs for C-/MAST, HC/PL*C-/MAST, and AX/PL*C-MAST interventions, with a total sample size of N = 320 participants.

## 2.4. Trial design and procedure

Participants completed two testing sessions (T1 and T2) one week apart, each lasting approximately 3.5 hours. To enhance the effectiveness of the stress intervention, C-/MAST exposure was conducted between 14:00 and 16:00, a period characterized by low endogenous CORT secretion (Trifonova et al., 2013). This trial followed a double-blind, factorial split-plot design (Kirk, 2013). As visualised in **Figure 3**, each participant completed both the MAST and C-MAST in counterbalanced order (within-participant factor) and was randomly assigned to receive either an atomoxetine (AX, 40 mg) or placebo (PL) capsule, as well as either a hydrocortisone (HC, 10 mg) or PL pill (between-participant factors). After AX/PL administration, participants completed questionnaires, practised both tasks from their randomly assigned Task Set (A or B), then received HC/PL and immediately completed the C-/MAST in a separate room, to prevent exogenous cortisol intake (HC) from suppressing stress-related cortisol secretion (Reuter, 2002; Spiga et al., 2014). Following C-/MAST, participants performed their assigned Task Set for another 60 minutes, divided into twelve 5-minute mini blocks. These mini-blocks were separated either by a 1-minute break or by saliva and mood sampling (every three blocks), as well as blood pressure, pulse measurement and adverse drug reaction assessment (four times per session), as visualised in **Figure 3C**.

### 2.5. Experimental interventions

#### 2.5.1. Maastricht Acute Stress Test (MAST) and its control test (C-MAST)

The standardised Maastricht Acute Stress Test (MAST) was implemented to induce acute stress responses, including the secretion of NE and CORT as well as other residual stress mediators (Quaedflieg et al., 2017; Smeets et al., 2012). Both the MAST and C-MAST was conducted by a female, opposite-sex experimenter, in a separate room, which was equipped with a water bath, two monitors (positioned to the left and behind the water bath), and a camera, and lasted for approximately 12 to 15 minutes (depending primarily on participants' reading speed). All C-/MAST instructions were presented on the left monitor. Transitions between the five hand immersion and four arithmetic phases, all of equal duration (though participants were unaware of the exact timing), were signalled by auditory cues and changed instructions on the left monitor. Throughout all C-/MASTs, participants and experimenters wore FFP2 masks.

The MAST consisted of alternating phases of hand immersions in ice-cold water (2–5°C) and a mental arithmetic task, where participants were required to count aloud, backward from 2043 in steps of 17. During both phases, participants' faces were recorded by the camera and projected onto the screen positioned behind the water bath. They were informed that their images would be captured for standardization purposes only because of the FFP2 mask, even if no actual recordings were made.

The C-MAST consisted of alternating phases of hand immersions in lukewarm water (38–41°C) and a simple mental arithmetic task, where participants were required to count aloud, backwards from 25 in single steps. Additionally, the camera and screen behind the water bath were switched off.

#### 2.5.2. Hydrocortisone (HC)

To evaluate whether CORT moderates the effects of acute stress on response accuracy and to distinguish its impact from other residual stress components, participants were randomly assigned to receive either 10mg of hydrocortisone (HC) or a placebo (PL; Mah et al., 2004; Miller et al., 2015; Weckesser et al., 2016). This dosage was chosen based on its excellent safety and tolerability record (Taylor et al., 2011), its established use in pharmacological treatment of adrenal insufficiency (Mah et al., 2004), as well as its similarity in magnitude to CORT increases induced by stress induction protocols (Kirschbaum et al., 1993). To minimize the potential suppression effects of HC on subsequent CORT secretion in response to the MAST, HC/PL was administered immediately before C-/MAST exposure (cf. **Figure 3C**; Weckesser et al., 2016).

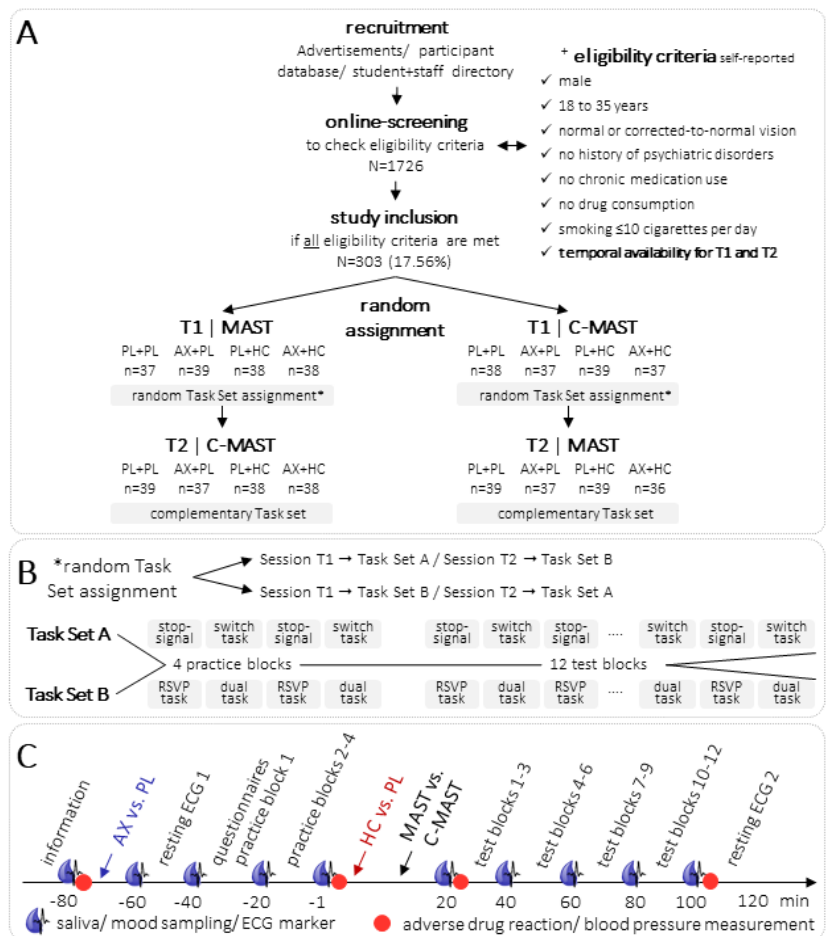

Figure 3. Schematic illustration of (A) the trial flow, including participant recruitment, online screening, and study inclusion for those meeting all eligibility criteria. Upon inclusion, participants randomly completed either the stress (MAST) or control test (C-MAST) at their first testing session (T1), and the alternate test at their second testing session (T2) one week later. Additionally, participants randomly received either 40 mg of atomoxetine (AX), 10 mg of hydrocortisone or placebos (PL). (B) To ensure a sufficient number of task trials, participants randomly completed either Task Set A, which included the stop-signal and switch tasks, or Task Set B, which included the rapid serial visual presentation (RSVP) and dual tasks at T1, with the remaining Task Set completed at T2. (C) Both testing sessions (T1 and T2) followed the same general procedure: Participants received AX or PL, completed questionnaires, practiced their assigned tasks with performance feedback, and then received HC or PL immediately before C-MAST/MAST exposure. This was followed by twelve test task blocks, presented in alternating order without performance feedback. Throughout each session, participants provided saliva and mood samples every 20 minutes and reported adverse drug reactions, blood pressure, and pulse measurements four times per session. Each session began and ended with a 15-minute recording of resting electrical cardiac activity.

### 2.5.3. Atomoxetine (AX)

To evaluate whether NE moderates the effects of acute stress on response accuracy and to distinguish its impact from other residual stress components, participants were randomly assigned to receive either 40mg of atomoxetine (AX), a selective NE reuptake inhibitor or a PL. This dosage was chosen based on its excellent safety and tolerability record, its pharmacokinetics (time to reach peak levels after oral intake), its prior use in studies on cognitive processing as well as its regulatory approval in Germany and the US for treatment of attention-deficit/hyperactivity disorder (Chamberlain et al., 2009; Sauer et al., 2005). To prolong NE availability after MAST exposure, AX was administered 80 minutes before C-/MAST exposure, as it requires approximately 60 to 180 minutes to reach peak plasma levels following oral administration (Sauer et al., 2005).

## 2.6. Saliva sampling and analysis

Saliva samples were collected using Salivette collection devices (Cortisol blue, Sarstedt, Nümbrecht, Germany) and frozen at −20°C after each session until analysis. Since HC is biochemically identical to CORT, its ingestion can contaminate the oral cavity, potentially leading to artificially elevated salivary cortisol concentrations that overestimate bioavailable CORT in the bloodstream (Perogamvros et al., 2010, 2011). To mitigate this, salivary cortisone, the enzymatically inactivated form of CORT, was analysed to indicate CORT availability (Miller & Plessow, 2013; Perogamvros et al., 2010). Additonally, α-amylase activity in saliva was measured to indicate NE availability (Rohleder et al., 2006). The saliva samples were analysed as follows: (1) Cortisone and cortisol concentrations were measured using liquid chromatography-tandem mass spectrometry, following the protocol outlined by Gao et al., (2015). (2) α-amylase activity was determined using an enzyme kinetic method, as described by Rohleder et al., (2006). In this manuscript, fourth-root transformed salivary cortisone and α-amylase activity served as a manipulation check, with further analyses specified in the preregistration record ISRCTN65723653 (https://doi.org/10.1186/ISRCTN65723653).

## 2.7. Self-report measures

Subjective mood was assessed using two short versions (presented in alternating order) of the German multidimensional mood questionnaire (*Multidimensionaler Befindlichkeitsfragebogen*, MDBF; Steyer et al., 1997). Participants rated their current experience of 12 adjectives related to good mood, calmness, and alertness (e.g., comfortable, relaxed, tired) on a 5-point Likert scale, ranging from "not at all" to "strongly". Adverse drug reactions were assessed using a short version of the German complaints list (*Beschwerden-Liste*; von Zerssen & Petermann, 2011). Participants rated the presence and intensity of 20 general physiological complaints (e.g., nausea, headache, chest pain) on a 4-point Likert scale, ranging from "not at all" to "strongly" (von Zerssen & Petermann, 2011).

After completing practice and test blocks of each task, participants rated mental demand using six subscales from the NASA Task Load Index (Hart & Staveland, 1988): [1] mental demand, [2] physical demand, [3] temporal demand, [4] performance self-evaluation, [5] motivation/effort and [6] frustration. These ratings were recorded on continuous visual analogue scales, anchored by the descriptors as "very low" to "very high" or "success" to "failure". The ratings were then aggregated into an overall indicator of perceived mental demand for task processing, ranging from 0% to 100% (Tsang & Velazquez, 1996).

## 2.8. Cardiac measurements

During each testing sessions, electrical cardiac activity was continuously recorded using portable 5-channel electrocardiogram (ECG) recorders (Medilog Darwin, Schiller, Germany). After a standardized instruction, blood pressure and pulse were measured independently by participants using arm cuffs (Omron Healthcare, Europe). Participants manually entered these values into an input mask on their test computer. Cardiac measurements are not presented in this manuscript.

## 2.9. Task design and technical setup

All tasks and stimuli were presented in a centered white serif font on a black background using the offline presentation tool of LABVANCED (version 1.0.71; Finger et al., 2017). Participants were seated approximately 50 cm from a 24-inch monitor. For all tasks (except the RSVP task), participants were instructed to respond as quickly and accurately as possible using a standard German Cherry QWERTZ keyboard. They were instructed to rest their fingers on designated and adjacent keys ("A" and "Y", "." and "Ö") to enter their responses. During practice blocks, performance feedback was provided for 100ms ("correct", "wrong", "too slow"). In both practice and test blocks, a prompt saying "please respond" appeared for 100ms if no response was made during four consecutive task trials. In addition to recording response accuracy (AC) and response times (RT), the number of completed trials in each 5-minute task block was assessed as a secondary outcome measure of participants' motivation. Each task block began with a brief reminder of the task rules and response mappings, followed by the prompt "please start the task by pressing the space key". If participants did not respond, the experiment was halted and could not proceed. If participants responded only after the 5 minutes expired, the next task block began, but no trails were recorded for the missed block (i.e., passive waiting did not enable participants to complete the experiment). If participants responded within the 5 minutes, they completed as many trials as possible at their own pace within the remaining time. This setup ensured active engagement while allowing accurate assessment of participants' performance and motivation.

### 2.9.1. Rapid serial visual presentation (RSVP) task

We adapted a rapid serial visual presentation (RSVP) task from Jolicœur (1999), in which participants identified two target digits within a sequence of 12 digits (see **Figure 4 A**). Each sequence contained both targets (T1 and T2): A yellow digit 2 or 6 (T1), was randomly presented at positions 2 through 10. A white digit 5 or 9 (T2), was always presented after T1 and randomly at positions 3 through 11. The remaining digits (0, 1, 3, 4, 7, and 8) served as white distractors, appearing in the first and last positions of each sequence as well as between T1 and T2. The number of distractors between T1 and T2 varied from 0 to 9, creating target stimulus onset asynchronies (SOA) between T1 and T2 ranging from 100ms to 1000ms. Participants first determined whether the yellow digit (T1) was a 6 by pressing "Y" for yes or "." for no. Next, they determined whether a 9 (T2) was presented afterwards using the same keys. Each trial began with a 250ms fixation cross, followed by the 12 digits presented at a rate of 100ms per digit. Participants had a maximum of 5000ms to respond to both T1 and T2. Each task block lasted 5 minutes and contained a variable number of trials.

The primary outcome measure was the correctness of T2 identifications ($AC_{T2C}$), conditional on the correct identification of T1 in each trial (which was averaged across participants to determine the overall probability of correct T2 detection). The secondary outcome measure was the number of completed trials.

### 2.9.2. Dual task

We adapted a dual task from Fischer and Hommel (2012), in which participants performed the same classification task twice. They had to indicate whether each of two digits was greater or smaller than 5 (see **Figure 4B**). The digits 2, 3, 7, and 8 were presented in task 1 (T1), while the digits 1, 4, 7, and 9 were presented in task 2 (T2). T1 and T2 digits appeared on either side of a fixation cross, with stimulus onset asynchronies (SOAs) of 80ms, 320ms, or 1280ms, equally distributed across trials. Participants were instructed to classify T1 first, pressing "Y" for digits smaller than 5 and "A" for digits greater than 5, and then to classify T2 using "." for digits smaller than 5 and "Ö" for digits greater than 5 (T1 priority). Each trial began with a 250ms fixation cross, followed by T1 and T2 displayed for 1000ms each. A blank screen was inserted if required by the SOA condition. Participants had a maximum of 1500ms to respond to both T1 and T2. Each task block lasted 5 minutes and contained a variable number of trials.

The primary outcome measure was the correctness of T2 classifications ($AC_{T2}$), which was averaged across participants to determine the overall probability of correct T2 classification. The secondary outcome measures were T2 response times (RT) and the total response time (sum of T1 and T2 RT) to have one measure that also includes T1 processing.

### 2.9.3. Stop signal task

We adapted a stop signal task from van der Schoot et al. (2005), in which participants classified digits as odd or even, unless a stop signal was presented, requiring them to inhibit their response (see **Figure 4C**). The stop signal, a red circle appearing around the digit, appeared in 25% of trials. The stop signal delay (SSD) was adjusted adaptively: It increased by 50ms after successful inhibition and decreased by 50ms after failed inhibition, within a range of 0ms to 1200ms. This adaptive design aimed to achieve a mean inhibition success rate of approximately 50%, as recommended by Verbruggen et al. (2019). Participants pressed "Y" to classify a digit as even and "." to classify it as odd. Each trial began with a 250ms fixation cross, followed by the digit, which was presented for 1000ms. If no stop signal appeared, the digit remained visible up to 1000ms and a black screen was inserted when required to reach a maximum response latency of 1500ms. Each task block lasted 5 minutes and contained a variable number of trials. The primary outcome measure was the stop signal delay (SSD) in each trial with a stop-signal. The secondary outcome measures were the estimated time needed for response inhibition (SS$_{RT}$), calculated using the integration method (Verbruggen et al., 2019), and the number of completed trials.

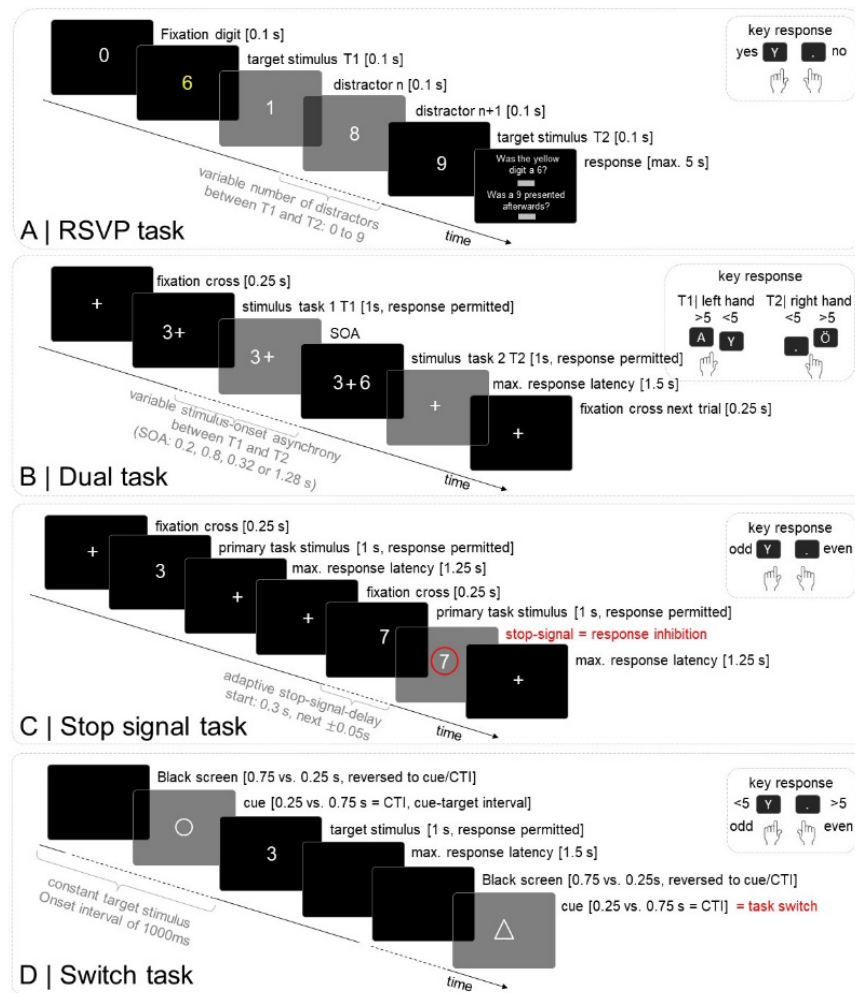

Figure 4. Schematic illustration of the four cognitive tasks used in this trial: (**A**) A rapid serial visual presentation (RSVP) task, (**B**) a dual task, (**C**) a stop-signal task and (**D**) a switch task. These tasks were selected to assess correctness of responses as primary outcome measure, indicating different cognitive processes. Specifically, (**A**) the RSVP task indicates low-demand, occipitally mediated visual-sensory processing, (**B**) the dual task indicates high-demand, frontally mediated cognitive flexibility to prioritize one of two concurrently presented tasks based on presentation order, (**C**) the stop-signal task indicates high-demand, frontally mediated response inhibition, and (**D**) the switch task indicates high-demand, frontally mediated cognitive flexibility to switch between two tasks based on a preceding visual cue. Each task was practised in two practice blocks followed by six test blocks, each lasting for 5 minutes. The total number of trials completed in each block served as a secondary outcome measure, along with response times.

### 2.9.4. Switch task

We adapted a switch task from Fischer and colleagues (2007), in which participants classified digits as greater or smaller than five or as odd or even (see **Figure 4 D**). A visual cue preceding the digit indicated the task: A triangle signalled the classification as greater or smaller than 5, while a circle signalled classification as odd or even. Both cues were presented in equal proportions. The cue preceded the digit (1 to 9, excluding 5) with either a long cue-target interval (CTI) of 750ms or a short CTI of 250ms, equally distributed. To maintain a constant trial length, the fixation cross at the beginning of each trial was presented for the opposing amount of time (250ms or 750ms). Participants had a maximum of 1500ms to classify each digit, which was presented for 1000ms. Each task block lasted 5 minutes and included a variable number of trials. The primary outcome measure was the correctness of digit classifications in each switch trial ($AC_{SWITCH}$). which was averaged across participants to determine the overall probability of correct digit classification. The secondary outcome measures were the response times (RT) to classify digits in switch trials and the number of completed task trials.

### 2.10. Statistical analysis

All statistical analyses were conducted using R (version 4.4.2), employing the blme package for Bayesian linear mixed-effects regression analyses, the emmeans package for estimating the marginal means of interest from the fitted models, and the ggplot2 and gridExtra packages for data visualization (Auguie, 2017; Chung et al., 2013; Lenth, 2025; Wickham, 2016).

### 2.10.1. Primary and secondary outcome measures

Across the four cognitive tasks of this trial, the correctness of digit classifications in test trials (0-80 minutes after C-/MAST cessation) were analysed as primary outcome (accuracy) measures:

– **RSVP task**: Probability of correct T2 detection when T1 was also correctly detected ($AC_{T2C}$)
– **Dual task**: Probability of correct T2 classification ($AC_{T2}$)
– **Switch task**: Probability of correct digit classification in switch trials ($AC_{SWITCH}$)
– **Stop signal task**: Probability of successful inhibition if a stop signal was presented ($AC_{STOP}$) as well as stop signal delay (SSD).

To provide a more comprehensive discussion of stress effects on cognitive processing and distinguish specific stress from rather general motivational effects, the following secondary and tertiary outcome measures were (exploratively) analysed:

– **Response times (RTs; secondary outcomes)**: Time in ms required for T2 classification ($RT_{T2}$) and both T1 and T2 classification (TRT) in the dual task (with the TRT indicating total response time or processing efficiency, with lower TRT indicating higher efficiency; Miller et al., 2009), for digit classification in switch trials ($RT_{SWITCH}$), or response inhibition ($SS_{RT}$). $SS_{RT}$ was estimated by the integration method as outlined by Verbruggen et al. (2019), including verification of the requirements listed there (which were met; see the online material provided). This involved replacing missing RTs for no-stop trials with the individual's maximum observed RT, the subtraction of mean stop signal delays from RTs at the cumulative RT distribution points that corresponds to the individual's probability of responding in stop trials.

- **Number of completed trials (N; tertiary outcomes)**: Average number of trials completed per participant and task.

### 2.10.2. Bayesian mixed-effect regression analysis of primary outcome measures

Bayesian mixed-effects regression analyses were conducted to evaluate the effects of acute stress and the moderating impact of AX and HC administration on primary outcome measures (correctness) across participants and sessions. Logistic regression models were specified incorporating random intercepts and slopes. The specified models accounted for task and trial design covariates (baseline model), the incremental impact of C-/MAST exposure and its interaction with time (stress model), and finally its interactions with AX and HC administration (interaction model). Default priors were used for inferring the covariance matrix of random effects. To control for error inflation, the gatekeeping principle was applied, meaning that the moderating impact of AX and HC (added in the interaction model) were only tested if C-/MAST exposure (added in the stress model) had a significant main effect at $\alpha = 0.05$ (Lehmacher et al., 1991).

Primary outcome measures were regressed on the conditional mean ($\beta_0$) and random deviations from $\beta_0$ for each participant p ($\alpha_{0p}$; random intercept). Fixed effects included testing session ($\beta_1$ SESSION: 0=first session, 1=second session) and order of C-/MAST exposure ($\beta_2$ ORDER: 0=C-MAST first, 1=MAST first), with random deviations from $\beta_1$ for each participant p being accounted for ($\alpha_{1p}$; random slope). Task-specific stimulus onset asynchronies (SOA) were included as categorical predictor encoded by i-1 dummy variables, accounting for variance due the number of distractors between T1 and T2 in the RSVP task, the delay between T1 and T2 in the dual task, and the interval between the cue and the target digit in the switch task, respectively (SOA). Time was also included as a categorical variable representing the contrast between early and late groupings of mini-task blocks (TIME). The **stress model (2)** extended the baseline model by adding acute stress exposure (MAST: 0=C-MAST, 1=MAST) and its interaction with time (MAST*TIME) to account for the proportion of variance induced by the stress intervention. The **interaction model (3)** further added the interactions between acute stress exposure and AX administration (MAST*AX) and acute stress exposure and HC administration (MAST*HC), along with their respective main effects (AX and HC), which were only included for exploratory purposes. Likelihood ratio tests and Nakagawa's conditional and marginal $R^2$ for mixed-effects models were used to compare the more parameterized models against the less parameterized ones. The relative likelihood of each of those three statistical models was calculated based on their small-sample Akaike Information Criteria (Wagenmakers & Farrell, 2004).

(1) **Baseline model**: Outcome= $(\beta_0 + \alpha_{0p}) + (\beta_1 + \alpha_{1p})$ SESSION + $\beta_2$ ORDER + $\beta_3$ TIME + $\beta_{4,...,(4+i-2)}$ SOA

(2) **Stress model**: Outcome= $(\beta_0 + \alpha_{0p}) + (\beta_1 + \alpha_{1p})$ SESSION + $\beta_2$ ORDER + $\beta_3$ TIME + $\beta_{4,...,(4+i-2)}$ SOA + $\beta_{4+i-1}$ MAST+ $\beta_{4+i}$ MAST*TIME

(3) **Interaction model**: Outcome= $(\beta_0 + \alpha_{0p}) + (\beta_1 + \alpha_{1p})$ SESSION + $\beta_2$ORDER + $\beta_3$ TIME + $\beta_{4,\ldots,(4+i-2)}$ SOA + $\beta_{4+i-1}$MAST+ $\beta_{4+i}$MAST*TIME + $\beta_{4+i+1}$MAST*AX+ $\beta_{4+i+2}$MAST*HC+ $\beta_{4+i+3}$AX + $\beta_{4+i+4}$HC

### 2.10.3. Mixed-effect regression analysis of salivary cortisol and amylase as manipulation check

To check the effectiveness of experimental interventions in inducing acute stress, or increasing CORT or NE availability, fourth root-transformed salivary cortisone levels, α-amylase activity, and subjective mood (good mood, calmness, alertness) were regressed as outcome variables on the same predictors that were included into the baseline, stress and interaction models (1) to (3), except for the tasks-specific (SOA) predictors. TIME was included as categorical variable representing the nine sampling time points every 20 minutes.

### 2.10.4. Data availability

All data used in these regression analyses, along with the task materials and analysis scripts, are publicly available at are publicly available at:
https://osf.io/6zyrb/?view_only=1ad0cb0630a5403b99662c09bcd819be

## 3    Results

### 3.1.    Manipulation check: Cortisone concentration, α-amylase activity and mood ratings

As visualized in **Figure 5**, the experimental interventions affected salivary cortisone concentrations, α-amylase activity and mood ratings as intended. Fixed and random effects of the statistical baseline model explained 23.01% and 25.23% of the variance in salivary cortisone concentrations, respectively, summing up to $R^2$=48.24%. MAST exposure and its interaction with time ($X^2(9)$= 31.84, p=0.0002; $\Delta R^2$=0.31%), as well as the ingestion of the pharmacological agents and their interactions with MAST exposure ($X^2(4)$=239.99, p<0.0001; $\Delta R^2$ =2.34%) relevantly affected the proportion of variance explained in salivary cortisol concentrations (cf. **Figure 5A**). Similarly, fixed and random effects of the statistical baseline model explained 4.75% and 42.68% of variance in salivary α-amylase activity, respectively, summing up to $R^2$=47.44%. MAST exposure and its interaction with time ($X^2(9)$= 20.32, p=0.0161; $\Delta R^2$=0.22%), as well as the ingestion of the pharmacological agents and their interactions with MAST exposure ($X^2(4)$=109.66, p<0.0001; $\Delta R^2$=0.44%) relevantly affected the proportion of variance explained in salivary α-amylase activity (cf. **Figure 5B**).

With respect to mood ratings, fixed and random effects of the baseline model explained 7.78% and 55.57% of the variance in good mood, respectively, summing up to $R^2$=64.12%. MAST exposure and its interaction with time ($X^2(9)$= 215.6, p<0.0001; $\Delta R^2$=1.43%), as well as the ingestion of the pharmacological agents and their interactions with MAST exposure ($X^2(4)$=14.77 p=0.0052; $\Delta R^2$=0.24%) relevantly affected the proportion of variance explained in good mood (cf. **Figure 5C**). Fixed and random effects of the baseline model explained 19.23% and 45.90% of the variance in calmness, respectively, summing up to $R^2$=65.13%. MAST exposure and its interaction with time ($X^2(9)$= 17.03, p=0.0483; $\Delta R^2$=0.08%), but not the ingestion of the pharmacological agents and their interactions with MAST exposure ($X^2(4)$=8.73, p=0.06812; $\Delta R^2$=1.6%) relevantly affected the proportion of variance explained in calmness (cf. **Figure 5D**). Fixed and random effects of the baseline model explained 6.62% and 48.83% of the variance in alertness, respectively, summing up to $R^2$=55.45%. MAST exposure and its interaction with time ($X^2(9)$=205.06, p<0.0001; $\Delta R^2$=1.63%), nor the ingestion of the pharmacological agents and their interactions with MAST exposure ($X^2(4)$=13.50, p=0.0009; $\Delta R^2$ =0.28%) displayed a relevant effect on the proportion of variance explained in alertness (cf. **Figure 5E**).

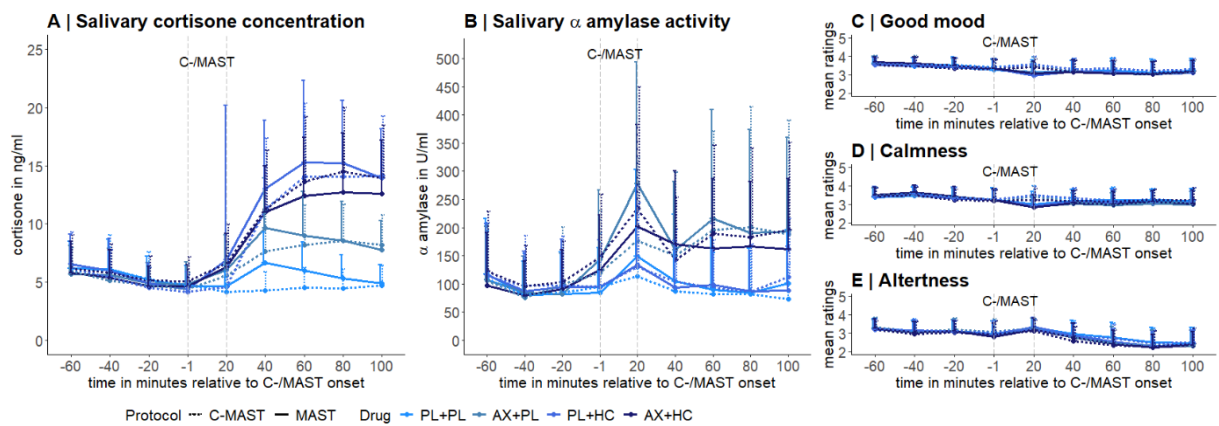

**Figure 5**. Mean and positive standard deviations of (**A**) salivary cortisone concentrations, (**B**) salivary α-amylase activities, or ratings of (**C**) good mood, (**D**) calmness and (**E**) alertness relative to C-/MAST onset at 0 min in all eight experimental conditions that resulted from crossing MAST and C-MAST exposure (within subject) with one of four pharmacological manipulations (between subject: placebo PL+PL, atomoxetine AX +PL, hydrocortisone HC+PL, or AX+HC). For clarity, only positive standard deviations are shown.

### 3.2. Manipulation check: Somatic complaints and substance detection

Neither MAST exposure and its interaction with time, nor the ingestion of the pharmacological agents and their interactions with MAST exposure, had a substantial effect on participants' reported levels of somatic complaints (all p's>0.1651; data provided online). Participants were generally unable to identify pharmacological manipulations beyond guessing probability (with the probability of receiving no drug being 25%, compared to 75% for receiving HC, AX or both; Prob[hit] = .58, p =1.0). Despite being informed about potential adverse reactions to each substance, participants seemed unable to identify which combination of substances they received (each combination PL+PL, AX+PL, HC+PL, AX+HC had an equal probability of Prob[hit]=0.25; p=0.5994).

### 3.3. Primary outcome measures
#### 3.3.1. Probability of correct T2 detection in the rapid serial visual presentation task

Participants correctly detected T2 when they have already correctly detected T1 ($AC_{T2C}$) with a mean probability of 66.03±47.36%. The fixed effects of the statistical baseline model explained $R^2$=3.99 % of variance in $AC_{T2C}$. Further 77.81% of variance in $AC_{T2C}$ were attributable to random effects, that is, systematic individual differences between participants. Neither MAST exposure and its interaction with time ($X^2(2)$=2.66, p=0.2651; $\Delta R^2$<0.01%), nor the ingestion of the pharmacological agents and their interactions with MAST exposure ($X^2(4)$=2.61, p=0.62; $\Delta R^2$<0.01%) displayed any relevant effect on the proportion of variance explained in $AC_{T2C}$ (see **Figure 6A** and **Table 1**). The baseline model yielded the largest relative likelihood (65.13%), whereas the stress (32.74%) and interaction models (2.13%) were less likely given the data.

### 3.3.2. Probability of correct T2 classification in the dual task

Participants correctly classified T2 ($AC_{T2}$) with a mean probability of 84.23±36.45%. The fixed effects of the baseline model explained $R^2$=5.77 % of variance in $AC_{T2}$. Further 90.55% of variance in $AC_{T2}$ were attributable to random effects, that is, systematic individual differences between participants. Neither MAST exposure and its interaction with time ($X^2(2)$=4.59, p=0.1009; $\Delta R^2$=0.01%), nor the ingestion of the pharmacological agents and their interactions with MAST exposure ($X^2(4)$=2.38, p=0.6654; $\Delta R^2$ <0.01%) displayed any relevant effect on the proportion of variance explained in $AC_{T2}$ (see **Figure 6B** and **Table 1**). The baseline and stress models yielded comparable relative likelihood (43.56% and 53.81%), whereas the interaction model (2.62%) was less likely given the data.

### 3.3.3. Probability of correct digit classification in switch trials of the switch task

Participants correctly classified digits with a mean probability of 90.42±29.44% in switch ($AC_{SWITCH}$) trials (M=407.37±101.97 of all task trials; as compared to a mean probability of correct responses of 93.52±24.61% in repetition trials). The fixed effects of the baseline model explained $R^2$=2.32% of variance in $AC_{SWITCH}$. Further 91.39% of variance in $AC_{SWITCH}$ were attributable to random effects, that is, systematic individual differences between participants. Neither MAST exposure and its interaction with time ($X^2(2)$=0.20, p=0.9063; $\Delta R^2$<0.01%), nor the ingestion of the pharmacological agents and their interactions with MAST exposure ($X^2(4)$=6.06, p=0.1945; $\Delta R^2$ <0.01%) displayed relevant effects on the proportion of variance explained in $AC_{SWITCH}$ (see **Figure 6C** and **Table 1**). The baseline model yielded the largest relative likelihood (83.67%), whereas the stress (12.10%) and interaction model (4.22%) were less likely given the data.

### 3.3.4. Probability of correct inhibition and stop-signal delay in the stop-signal task

Participants successfully inhibited their response (AC$_{STOP}$) on 49.02 ± 6.60% of stop-signal trials (M = 286.06 ± 50.43 trials), with a mean stop-signal delay of 433.66 ± 230.30 ms. This mean stopping probability slightly deviates slightly from the programmed (intended) 50% due to variability in the number of trials completed across participants (Verbruggen et al., 2019).

The fixed effects of the baseline model explained R²= 0.22% of variance in AC$_{STOP}$. Further 46.03% of variance in AC$_{STOP}$ were attributable to random effects, that is, systematic individual differences between participants. MAST exposure and its interaction with time ($X^2(2)$=293.78, p<0.0001; ΔR²=0.06%), but not the ingestion of the pharmacological agents and their interactions with MAST exposure ($X^2(4)$=2.56, p=0.6339; ΔR² =-0.05%) significantly affected the proportion of variance explained in AC$_{STOP}$. As listed in **Table 1**, MAST exposure slightly decreased AC$_{STOP}$ later in time (($β_{MAST}$=0.02±0.01; $β_{TIME}$=-0.001±0.001, $β_{MAST*TIME}$=-0.2±0.001) and when HC was co-administered ($β_{MAST*HC}$=-0.01±0.01). The stress model yielded the largest relative likelihood (~100%), whereas the baseline (0.00%) and interaction model (0.00%) were extremely unlikely given the data.

The fixed effects of the baseline model explained R²=2.77 % of variance in SSD. Further 68.43% of variance in SDD were attributable to random effects, that is, systematic individual differences between participants. MAST exposure and its interaction with time ($X^2(2)$=34.20, p<0.0001; ΔR²=0.08%), but not the ingestion of the pharmacological agents and their interactions with MAST exposure ($X^2(4)$=7.16, p=0.1275; ΔR² =0.22%) significantly affected the proportion of variance explained in SSD. As visualised in **Figure 6D** and listed in **Table 1**, MAST exposure ($β_{MAST}$=-28.96±38.99) slightly reduced SDD especially later in time ($β_{TIME}$=-40.85±1.27, $β_{MAST*TIME}$=10.45±1.80) and when AX was co-administered ($β_{MAST*AX}$=-69.83±45.05). The interaction model yielded the largest relative likelihood (~100%), whereas the baseline (0.00%) and stress model (0.00%) were extremely unlikely given the data.

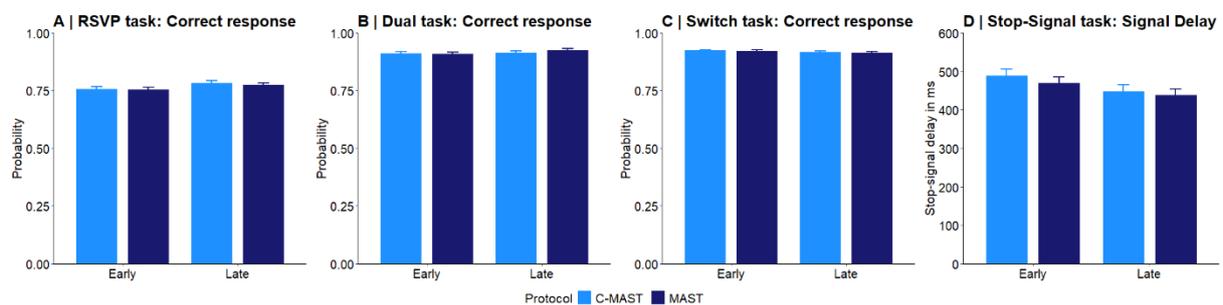

Figure 6. Estimated marginal probability of correct responses (with standard error of the mean, SE) across the (A) rapid visual serial presentation (RSVP) task, (B) dual task, (C) switch trials of the switch task or (D) stop-signal delay in the stop signal task. General notes: ■ indicates performance after the Maastricht Acute Stress Test (MAST) or ■ its control test (C-MAST); Early and late phases refer to the time elapsed since the end of the C-/MAST, with the early phase (0-30 minutes post-offset) dominated by stress-induced norepinephrine (NE) availability and the late phase (30-60 minutes post-offset) dominated by cortisol (CORT) availability.

### 3.4. Secondary outcome measures

### 3.4.1. Response time for T2 classification in the dual task

On average, participants needed 719.51±224.73ms to classify T2 ($RT_{T2}$) in the dual task. The fixed effects of the baseline model explained $R^2$=25.31 % of variance in $RT_{T2}$. Further 19.38% of variance in $RT_{T2}$ were attributable to random effects, that is, systematic individual differences between participants. MAST exposure and its interaction with time ($X^2(2)$=7.16, p=0.0279; $\Delta R^2$=0.0%), but not the ingestion of the pharmacological agents and their interactions with MAST exposure ($X^2(4)$=7.92, p=0.0943; $\Delta R^2$ =0.1%) relevantly affected the proportion of variance explained in $RT_{T2}$. As visualised in **Figure 7A** and listed in **Table 1**, $RT_{T2}$ decreases later in time ($β_{TIME}$=-23.60±1.45) but slightly less so after MAST exposure ($β_{MAST}$=-1.87±11.60, $β_{MAST*TIME}$=5.48±2.07). The interaction model yielded the largest relative likelihood (99.99%), whereas the baseline (0.00%) and stress model (0.00%) were unlikely given the data.

### 3.4.2. Total response time for T1 and T2 classification in the dual task

On average, participants needed 1498.07±459.87ms to classify both T1 and T2 (TRT) in the dual task. The fixed effects of the statistical baseline model explained $R^2$=4.17 % of variance in TRT. Further 29.77% of variance in TRT were attributable to random effects, that is, systematic individual differences between participants. MAST exposure and its interaction with time ($X^2(2)$=16.52, p<0.0001; $\Delta R^2$=0.1%), but not the ingestion of the pharmacological agents and their interactions with MAST exposure ($X^2(4)$=9.15, p=0.0058; $\Delta R^2$ =0.2%) relevantly affected the proportion of variance explained in TRT. As visualised in **Figure 7B** and listed in **Table 1**, TRT decreases later in time ($β_{TIME}$=-57.04±3.23), even more after concurrent MAST exposure ($β_{MAST}$=-23.14±29.29, $β_{MAST*TIME}$=18.68±4.60). The stress model yielded the largest relative likelihood (99.99%), whereas the baseline (0.00%) and interaction model (0.00%) were unlikely given the data.

### 3.5. Response time of digit classification in the switch task

On average, participants needed 646.84±195.13ms to classify digits in the switch trials ($RT_{SWITCH}$) of the switch task (as compared to 619.15±178.79ms in repeat trials). The fixed effects of the baseline model explained $R^2$=4.39 % of variance in $RT_{SWITCH}$. Further 20.59% of variance in $RT_{SWITCH}$ were attributable to random effects, that is, systematic individual differences between participants. Neither MAST exposure and its interaction with time ($X^2(2)$=0.20, p=0.9063; $\Delta R^2$=0.1%), nor the ingestion of the pharmacological agents and their interactions with MAST exposure ($X^2(4)$=6.06, p=0.1945; $\Delta R^2$ =0.3%) displayed any relevant effect on the proportion of variance explained in $RT_{SWITCH}$ (see **Figure 7C** and **Table 1**). The baseline model yielded the largest relative likelihood (83.67%), whereas the stress (12.10%) and interaction model (4.22%) were less likely given the data.

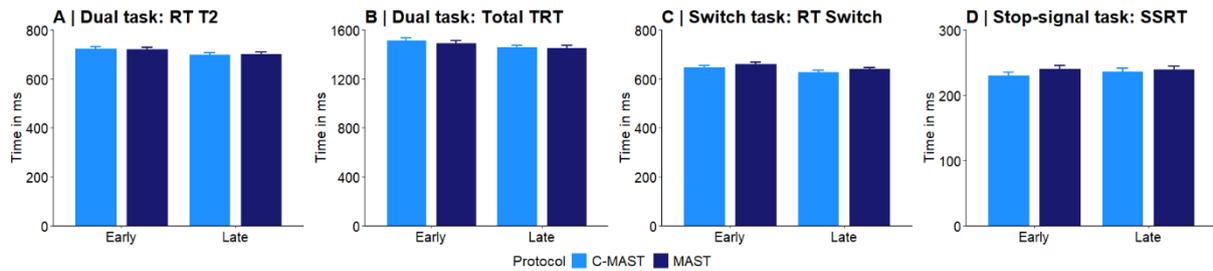

Figure 7. Estimated marginal time (with standard error of the mean, SE) to respond to (A) the second target (task) in the dual task, (B) both targets (tasks) in the dual task, (C) digits in switch trials of the switch task or estimated time needed to (D) inhibit the response in the stop signal task. General notes: ■ indicates performance after the Maastricht Acute Stress Test (MAST) or ■ its control test (C-MAST); Early and late phases refer to the time elapsed since the end of the C-/MAST, with the early phase (0-30 minutes post-offset) dominated by stress-induced norepinephrine (NE) availability and the late phase (30-60 minutes post-offset) dominated by cortisol (CORT) availability.

### 3.5.1. Response inhibition time in the stop signal task

On average, participants required 236.76±77.85ms to inhibit their response on trials in which a stop signal was presented ($SS_{RT}$). The fixed effects of the baseline model explained $R^2$=0.64 % of variance in $SS_{RT}$. 62.01% of variance in $SS_{RT}$ were attributable to random effects, that is, systematic individual differences between participants. MAST exposure and its interaction with time ($X^2(2)$=147.01, p<0.0001; $\Delta R^2$=0.1%), but not the ingestion of the pharmacological agents and their interactions with MAST exposure ($X^2(4)$=5.50, p=0.2393; $\Delta R^2$=0.3%) relevantly affected the proportion of variance explained in $SS_{RT}$. As visualised in **Figure 7D** and listed in **Table 1**, MAST exposure increased SSRT ($\beta_{MAST}$=17.54±19.88) especially early in time ($\beta_{TIME}$=-23.60±1.27, $\beta_{MAST*TIME}$=5.49±2.01). The interaction model yielded the largest relative likelihood (99.99%), whereas the baseline (0.00%) and interaction model (0.00%) were unlikely given the data.

### 3.5.2. Number of completed task trials

During the 60-minute period following C-/MAST exposure, participants completed an average of 493.55±51.51 trials in the RSVP task, 732.01±66.51 trials in the dual task, 826.60±60.01 in the switch task (including repeat trials) and 1127.33±122.85 trials in the stop-signal task (including trials without a stop-signal) per session. Results of the analyses of these numbers of completed task trials are presented in **Table 1**. Neither MAST exposure and its interaction with time ($X^2(2)$=3.48, p=0.1758; $\Delta R^2$=0.0%), nor the ingestion of the pharmacological agents and their interactions with MAST exposure ($X^2(4)$=6.49, p=0.1657; $\Delta R^2$=0.1%) displayed relevant effects on the proportion of variance explained in the number of completed RSVP task trials. Similarly, neither MAST exposure and its interaction with time ($X^2(2)$=3.73, p=0.1553; $\Delta R^2$=0.0%), nor the ingestion of the pharmacological agents and their interactions with MAST exposure ($X^2(4)$=6.19, p=0.1854; $\Delta R^2$=0.2%) displayed relevant effects on the proportion of variance explained in the number of completed dual task trials. Comparably, neither MAST exposure and its interaction with time ($X^2(2)=1.44$, $p=0.4867$; $\Delta R^2=0.1\%$), nor the ingestion of the pharmacological agents and their interactions with MAST exposure ($X^2(4)=4.67$, $p=0.3226$; $\Delta R^2 =0.4\%$) displayed relevant effects on the proportion of variance explained in the number of completed stop-signal task trials. In contrast, however, MAST exposure and its interaction with time ($X^2(2)= 9.14$, $p=0.0104$; $\Delta R^2=1.2\%$), but not the ingestion of the pharmacological agents and their interactions with MAST exposure ($X^2(4)=9.01$, $p=0.0609$; $\Delta R^2 =1.0\%$) displayed relevant effects on the proportion of variance explained in the number of completed trials in the switch task. Specifically, MAST exposure reduced the number of completed trials, especially later in time ($\beta_{TIME}=-14.83\pm3.76$; $\beta_{MAST}==-0.38\pm3.87$; $\beta_{MAST*TIME}=-11.25\pm5.32$; cf. **Table 1**).

## 4 Discussion

This randomised controlled trial assessed cognitive task performances under experimentally induced acute stress and pharmacologically increased stress hormone availability to rigorously test which of three influential cognitive stress effect models best predicts the mechanisms, scope and temporal dynamic of stress-related effects. Across 606 testing sessions, the Maastricht Acute Stress Test (MAST) compared to its control test reliably increased salivary cortisone and α-amylase activity over time, reduced good mood and calmness, and increased alertness ratings, indicating a successful implementation of complex acute stress responses. With respect to cognitive performance under stress, MAST exposure consistently impaired response inhibition in the stop signal task, as indicated shorter stop-signal delays (SSD), lower probabilities of successful stopping ($AC_{STOP}$), and prolonged stop-signal reaction times ($RT_{STOP}$), particularly during later testing phases (40–80 minutes post-stress). By contrast, MAST exposure stabilized or protected cognitive processing from time-related decline in the dual task, as reflected by reduced total response times (TRTs; indexing processing efficiency) but increased time to classify the second of both tasks ($RT_{T2}$; indexing less successful prioritization or task shielding in the favour of higher processing efficiency). Performance in the remaining RSVP and switch tasks remained largely unaffected. Pharmacological manipulations produced no somatic complaints and could not be identified better than chance. Although atomoxetine (AX) and hydrocortisone (HC) selectively increased salivary α-amylase (indicating noradrenergic activity) and cortisone (indicating cortisol availability), confirming the effectiveness of the manipulations, no model-consistent interaction effects (with MAST exposure) were observed. Overall, the observed pattern of task-specific impairment alongside stabilized cognitive performance under acute stress cannot be fully explained by any of the three tested stress-effect models (see **Figure 2** for their predicted performance changes under stress). Although this trial provides one rigorous empirical evaluation of competing stress-effect models within a single experimental trial, its findings could not advance the mechanistic understanding of how acute stress affects human cognition in the desired amount or way. Rather, the observed pattern of results highlights the need to refine existing models and adopt more integrative approaches to advance a mechanistic understanding of acute stress effects on cognitive task performance in laboratory and prospectively of course, in real-world contexts.

A notable finding regarding the effect of stress on response inhibition is that, in the present study, stress exerted a negative effect, whereas the most influential meta-analysis to date reports an overall improvement in response inhibition under acute stress (Shields et al., 2016) and HC exposure (Shields et al., 2015). Although both meta-analyses included primary studies reporting negative effects, those most closely resembling the present design with respect to stressor type, timing, and task consistently observed stress-related improvements in response inhibition (Cackowski et al., 2014; Schwabe et al., 2013). More recent, comparable studies reporting absent behavioural stress effects likewise offer limited insight into potential moderators underlying this discrepancy (Dierolf et al., 2017; Jiang & Rau, 2017). Although one hypothesis is that parasympathetic activation mediates stress-related improvements in response inhibition (Roos et al., 2017), this would also predict a concomitant negative effect of atomoxetine, given its sympathomimetic properties, which was not observed in this trial. Thus, although we detected a stress-induced impairment in response inhibition, this finding provides limited leverage for advancing a mechanistic understanding of how acute stress modulates inhibitory control processes.

We observed a divergent stress effect pattern on dual-task performance, comprising of improved processing efficiency (TRT) alongside reduced task prioritization or shielding ($RT_{T2}$), particularly later in time when only cortisol remains elevated following laboratory stress-induction (see also **Figure 5**). The stress-induced impairment ($RT_{T2}$) is consistent with one of our earlier studies employing a comparable design (Weckesser et al., 2016). However, in contrast to this earlier study, we were unable to replicate the positive HC effect on dual-task performance ($RT_{T2}$), or its complete mediation via visual perception (Miller et al., 2015; Weckesser et al., 2016; Schwabe & Wolf, 2010). Consequently, our earlier computationally informed account that cortisol (mimicked by HC) might counteract adverse (residual) effects of acute stress on executive processes (as indicated by $RT_{T2}$) by improving visual perception (most closely associated with RSVP performance $AC_{T2C}$), is not supported in the present trial and the absence of HC effects alongside a preserved RSVP performance ($AC_{T2C}$) under both MAST and HC exposure (Weckesser et al., 2016).

There are several general limitations of this trial that are worth to be mentioned. First, the sample was relatively homogenous in terms of gender (only men) and age (18-35 years) and was recruited primarily through a university mailing list (i.e., it consisted primarily of students), a participant database, and local advertisements in Dresden. This recruitment strategy may have resulted in a reduction of residual variance in cognitive performance, but may also have resulted in a comparatively high-funtioning sample, whose cognitive performance may be more resilient to short-term disruptions by laboratory stressors or singular pharmacological manipulation of stress hormone availbaility (Fan, & Yin, 2001; Novotný et al., 2021). Related to this, the data collection of this trial began during the COVID-19 pandemic, requiring participants to wear FFP2 masks during the C-MAST and MAST procedures, which may have reduced differences between the stress and control condition (although they were present and statistically meaningful; see Figure 5) and/or subsequent cognitive task performance. Comparably, the selected AX dose might have been too low to sufficiently increase the magnitude and duration of NE availability from becoming behaviourally effective and thus inducing changes in the task performance measures we assessed in our high-functional student sample without self-reported mental disorders (Chamberlain et al., 2006, 2007; Hernaus et al., 2018; Michelson et al., 2003; Tona et al., 2020). Furthermore, to ensure a sufficient number of trials per task, we presented only two of the four tasks of interest, with a variable number of task trials per participant. While this decision was methodologically justified—considering trials as motivational indicators, the timing of saliva samples, and the reliability of performance estimates—it, like any design choice, could have contributed to non-specific variance in our data.

Table 1. Summary of Bayesian mixed-effect regression analysis of primary and secondary outcome measures. Outcomes measures testing session ($\beta_1$SESSION: 1st or 2nd session), order of C-/MAST exposure ($\beta_2$ORDER: whether C-/MAST was administered first), their interaction ($\beta_3$ORDERSESSION), task-specific stimulus onset asynchrony (SOA) ($\beta_4$SOA), and time across 12 mini-task blocks ($\beta_5$TIME) in the baseline model (not shown here). Additional predictors included exposure to the Maastricht Acute Stress Test (MAST) or its control condition, its interaction with time (MAST*TIME) in the stress model, as well as its interaction with atomoxetine (MAST*AX) or hydrocortisone (MAST*HC), alongside the main effects of atomoxetine (AX) and hydrocortisone (HC) in the interaction model. Primary outcome measures were the probability of correct responses in the rapid-serial visual presentation (RSVP) task ($AC_{T2C}$) when the first target was already correctly identified, dual ($AC_{T2}$), switch ($AC_{SWITCH}$) and stop-signal ($AC_{STOP}$) tasks, along with the stop-signal delay (SSD). Secondary outcome measures included response times (RTs) for classifying the prioritized second digit ($RT_{T2}$) and total response time (TRT) in the dual task, response times in switch trials ($RT_{SWITCH}$) of the switch task, and estimated stop-signal reaction times ($SS_{RT}$). Tertiary outcomes were the number of completed task trials (N) across all four tasks ($N_{RSVP}$, $N_{DUAL}$, $N_{SWITCH}$, $N_{SS}$). Models allowed for random variation in intercepts and slopes for SESSION. $R^2$ values represent the estimated proportion of explained variance, incorporating both fixed and random effects.

| | | Baseline model | stress model | | | | | interaction model | | | | | | | |
|---|---|---|---|---|---|---|---|---|---|---|---|---|---|---|---|
| Task | Outcome | $R^2$ % | β INTERCEPT | β TIME | β MAST | β MAST*TIME | $R^2$ % | β INTERCEPT | β TIME | β MAST | β MAST*TIME | β MAST*AX | β MAST*HC | β AX | β HC | $R^2$ |
| RSVP | $AC_{T2C}$ | 81.80 | 0.99±0.08 | 0.15±0.02 | -0.01±0.08 | -0.04±0.03 | 81.80 | 1.03±0.11 | 0.15±0.02 | -0.15±0.13 | -0.04±0.03 | 0.10±0.15 | 0.20±0.15 | 0.00±0.11 | -0.10±0.11 | 81.80 |
| Dual | $AC_{T2}$ | 96.32 | 2.01±0.18 | 0.02±0.06 | -0.03±0.15 | 0.19±0.09 | 96.31 | 2.06±0.23 | 0.02±0.06 | -0.18±0.26 | 0.19±0.09 | 0.28±0.29 | 0.02±0.29 | 0.00±0.21 | -0.11±0.21 | 96.31 |
| Switch | $AC_{SWITCH}$ | 93.72 | 2.65±0.09 | -0.09±0.03 | -0.02±0.09 | -0.01±0.04 | 93.72 | 2.45±0.12 | -0.10±0.03 | 0.18±0.15 | -0.01±0.04 | -0.13±0.17 | -0.28±0.17 | 0.14±0.12 | 0.25±0.12 | 93.72 |
| Stop-signal | $AC_{STOP}$ | 47.99 | 0.49±0.005 | -0.001±0.001 | 0.001±0.01 | -0.02±0.001 | 46.51 | 0.49±0.01 | -0.001±0.001 | 0.002±0.009 | -0.02±0.001 | 0.007±0.010 | -0.008±0.010 | -0.00±0.007 | 0.01±0.007 | 46.04 |
| Stop-signal | SSD | 71.19 | 450.52±21.92 | -40.85±1.27 | -20.04±22.40 | 10.45±1.80 | 71.27 | 461.90±31.13 | -40.85±1.27 | -26.66±38.73 | 10.45±1.80 | -73.98±44.52 | 86.33±44.55 | 38.23±31.64 | -61.28±31.65 | 71.49 |
| Dual | $RT_{T2}$ | 44.69 | 883.85±11.22 | -23.60±1.45 | -1.87±11.59 | 5.48±2.07 | 44.73 | 867.37±16.18 | -23.60±1.45 | 17.53±19.88 | 5.49±2.07 | -20.19±22.93 | -18.58±22.98 | -5.19±16.18 | 36.90±16.18 | 44.86 |
| Dual | TRT | 33.94 | 1672.04±31.50 | -57.04±3.23 | -23.14±29.29 | 18.68±4.60 | 34.03 | 1618.67±42.65 | -57.05±3.23 | 31.96±49.67 | 18.69±4.60 | -44.69±57.14 | -65.20±57.25 | -9.31±40.34 | 112.29±40.35 | 34.22 |
| Switch | $RT_{SWITCH}$ | 24.99 | 600.90±8.70 | -20.64±1.43 | 14.11±10.43 | 0.19±2.03 | 24.88 | 609.60±13.22 | -20.64±1.43 | 6.13±17.54 | 0.19±2.03 | 18.89±19.77 | -2.13±19.79 | -17.41±14.07 | -0.13±14.08 | 25.21 |
| Stop-Signal | $SS_{RT}$ | 62.64 | 230.12±6.37 | 6.49±0.47 | 13.81±5.53 | -7.92±0.66 | 62.76 | 236.22±8.95 | 6.49±0.47 | 0.91±11.81 | -7.92±0.66 | 28.34±12.27 | -9.86±12.41 | -16.31±9.42 | 3.97±9.47 | 63.08 |
| RSVP | $N_{RSVP}$ | 82.82 | 79.60±0.74 | 3.11±0.32 | -1.20±0.74 | -0.28±0.45 | 82.83 | 80.28±1.03 | 3.11±0.32 | -2.78±1.24 | -0.28±0.45 | 0.99±1.40 | 2.17±1.41 | 0.46±0.99 | -1.77±0.99 | 83.01 |
| Dual | $N_{DUAL}$ | 73.65 | 128.10±3.40 | -99.41±3.61 | 0.003±3.78 | -7.24±5.13 | 73.75 | 123.82±4.39 | -99.41±3.62 | 5.46±5.41 | -7.24±5.13 | -0.92±5.50 | -10.02±5.50 | -0.79±3.88 | 9.10±3.88 | 73.72 |
| Switch | $N_{SWITCH}$ | 13.65 | 147.37±3.30 | -14.83±3.76 | -0.38±3.87 | -11.25±5.32 | 14.84 | 150.51±4.30 | -14.83±3.75 | -7.24±5.50 | -11.25±5.30 | 4.23±5.57 | 9.43±5.57 | 2.53±3.94 | -8.90±3.94 | 15.81 |
| Stop-signal | $N_{SS}$ | 60.18 | 173.42±2.54 | -11.21±1.89 | 0.67±2.93 | 2.56±2.68 | 60.29 | 169.03±3.59 | -11.21±1.89 | 5.86±4.66 | 2.56±2.68 | -3.36±5.12 | -6.99±5.12 | 1.67±3.62 | 7.23±3.63 | 60.70 |

Note. Coefficient estimates for SOA and random effects are not shown for reasons of clarity.

**Authors contributions:**

– Lisa Weckesser: Funding acquisition, conceptualization, project administration, investigation, Formal analysis, Software, Visualization, Writing of the original draft;

– Charlotte Grosskopf: Investigation, writing - review & editing;

– Benjamin Weber: Investigation, writing - review & editing;

– Selen Soylu: Writing - review & editing;

– Tanja Endrass: Writing - review & editing;

– Robert Miller: Funding acquisition, conceptualization, methodology, supervision, review & editing;

**Declaration of generative AI and AI-assisted technologies in the writing process:** In preparing this manuscript, LW used ChatGPT 4.0 to enhance language and readability by requesting, "Please improve the language and readability of the following paragraph". The author reviewed and modified the AI-generated content as needed and takes full responsibility for the final content of this manuscript.